\documentclass[aps,prd,letterpaper,twocolumn,superscriptaddress,showpacs]{revtex4}

\usepackage{graphicx} 
\usepackage{multirow}
\usepackage{wasysym} 

\DeclareTextSymbol{\degrees}{OT1}{23}

\begin{document}


\title{Continuum gamma-ray emission from light dark matter\\ positrons and electrons}


\author{P.~Sizun}
\email[]{sizun@cea.fr}
\affiliation{DAPNIA/Service d'Astrophysique, CEA~Saclay, F-91191 Gif-sur-Yvette Cedex, France}
\author{M.~Cass\'e}
\affiliation{DAPNIA/Service d'Astrophysique, CEA~Saclay, F-91191 Gif-sur-Yvette Cedex, France}
\affiliation{Institut d'Astrophysique de Paris, 98~bis Bd Arago, F-75014 Paris, France}
\author{S.~Schanne}
\affiliation{DAPNIA/Service d'Astrophysique, CEA~Saclay, F-91191 Gif-sur-Yvette Cedex, France}


\date{\today}

\begin{abstract}
   The annihilation of light dark matter was recently advocated as a possible explanation of the large positron injection rate at the Galactic center deduced from observations by the SPI spectrometer aboard INTEGRAL. The modelling of internal Bremsstrahlung and in-flight annihilation radiations associated to this process drastically reduced the mass range of this light dark matter particle.
	We estimate critically the various energy losses and radiations involved in the propagation of the positron before its annihilation ---\,in-flight or at rest.

   Using a simple model with mono-energetic positrons injected and confined to the Galactic bulge, we compute energy losses and gamma-ray radiations caused by ionization, Bremsstrahlung interactions as well as in-flight and at rest annihilation and compare these predictions to the available observations, for various injection energies.

   Confronting the predictions with observations by the CGRO/EGRET, CGRO/COMPTEL, INTEGRAL/SPI and INTEGRAL/IBIS/ISGRI instruments, we deduce a mass upper bound of 3 to 7.5~$\mathrm{MeV/c^2}$ for the hypothetical light dark matter particle. The most restrictive limit is in agreement with the value previously found by  Beacom and Y\"{u}ksel and was obtained under similar assumptions, while the 7.5~$\mathrm{MeV/c^2}$ value corresponds to more conservative choices and to a partially ionized propagation medium. We stress how the limit depends on the degree of ionization of the propagation medium and how its precision could be improved by a better appraisal of data uncertainties.
\end{abstract}


\maketitle


\section{Introduction}
Observed by various instruments since the 70's \cite{Milne2006}, the 511~keV gamma-ray line emission from the Galactic center has now been studied for almost four years with the SPI spectrometer \cite{Vedrenne2003} aboard the INTEGRAL satellite. While this emission is definitely the result of electron-positron annihilations in the central region of the Galaxy, the origin of these positrons remains obscure.

What is clear, however, from the spectral analysis of this line, is that most of the positrons are set to rest, form positronium and annihilate in the Galactic bulge (hereafter GB) \cite{Knodlseder2005,Jean2006,Churazov2005}. Thus, there should be material in the bulge to slow down these leptons, as well as a significant magnetic field to confine them, which is an important information since the interstellar medium in this region is very difficult to observe. As far as positrons are concerned, the main characteristic to explain is their very high injection rate, \emph{i.e.} $\sim{}1.4\times{}10^{43}~\mathrm{e^{+}s^{-1}}$, as deduced from the 511~keV line flux of $(1.07\pm{}0.03)\times{}10^{-3}~\mathrm{ph\,cm^{-2}\,s^{-1}}$ measured by INTEGRAL/SPI \cite{Jean2006}.
Though previous instruments like SMM, TGRS , CGRO/OSSE gave essentially the same requirement \cite{Milne2006}, it is surprising that nobody had remarked that type Ia supernov\ae{}, a long time considered the main source of GB positrons, fall short explaining this high injection rate \cite{Casse2004,Schanne2004,Schanne2004a}. Fortunately, this drawback is now taken seriously.

Another clue to the origin of these GB positrons is the large extent of the emitting region, 8\degrees{} FWHM \cite{Knodlseder2005}, indicating that either they emanate from a single central source or a population of sources densely packed in the center of the Galaxy and diffuse to fill the whole bulge region, or there is a source population with an extension similar or greater than the GB. Indeed, the 511~keV emission region traces where positrons annihilate, and is therefore related to the distribution of the gas in the Galactic bulge.

Potential astrophysical sources of positrons are numerous (SNIa, hypernov\ae{}, gamma-ray bursts, cosmic rays, low mass X-ray binaries, millisecond pulsars,\ldots) and we will not discuss them in detail here. None of them qualifies to account for both the flux and morphology of the 511~keV emission.
Some candidates are excluded on the basis of the disk over bulge ratio \cite{Prantzos2006} and others like millisecond pulsars \cite{Wang2005} and gamma-ray bursts \cite{Parizot2005,Bertone2006} are shaded off, the first because the electrons and positrons would be injected at such energies that their Bremsstrahlung emission would be excessive, and the second because their rate of appearance is insufficient to supply positrons at the required rate \cite{Woosley2006,Fruchter2006,Stanek2006}.

  
In front of this situation, two rather daring hypotheses have been put forward~:
\begin{itemize}
	\item[-] the sources belong to the established list but there is something important not taken into account in the astrophysical scenario. Positrons from the Galactic disk can be transferred to the bulge where they annihilate \cite{Prantzos2006}.
However, the transport mechanism from the disk to the center is poorly known, depending critically on the topology and strength of the large scale magnetic field and on the diffusion coefficient of low energy positrons in the relevant magnetic field \cite{Parizot2005,Jean2006};
\item[-] new sources or mechanisms are required, which are not in the catalog. We will consider this second proposal, and more specifically the light dark matter (hereafter LDM) scenario \cite{Boehm2004a,Casse2004,Fayet2006,Casse2006,Rasera2006,Schanne2004}. Other exotica from the particle physics community include Q balls \cite{Kasuya2005}, relic particles \cite{Picciotto2005}, decaying axinos \cite{Hooper2004}, primordial black holes \cite{Frampton2005}, color superconducting dark matter \cite{Oaknin2005}, superconducting cosmic strings \cite{Ferrer2005}, dark energy stars \cite{Chapline2005}, and moduli decays \cite{Kawasaki2005}; they will not be discussed  here.
\end{itemize}

Surprisingly, though the 511 keV emission from the GB was first detected more than thirty years ago, it is only with the SPI observations that a great deal of imagination and activity has been triggered among the astroparticle community.

In this contribution, we gather $\gamma$-ray observations performed by the CGRO (COMPTEL, EGRET) and INTEGRAL (IBIS/ISGRI, SPI) missions and see what constraints they set on the LDM candidate. A note of warning however: since we are interested in diffuse radiation, the delicate problems of unresolved sources, particularly acute at low energies, and of alternate diffuse sources should always be kept in mind.

We do not focus on the theoretical morphology of the LDM emission \cite{Boehm2004c,Rasera2006}, which is a thorny subject. It first depends on the dark matter distribution, which is not well established, varies considerably from one author to the other \citep[see \emph{e.g.}][]{Navarro1997,Moore1999} and could furthermore be clumpy. It also depends on the complex spatial propagation of the leptons resulting from LDM annihilation and on the distribution of the ISM matter with which they interact, also poorly known. Therefore, we use the morphological information we dispose of, \emph{i.e.} that of the radiation from positron annihilations occurring at rest, as observed by INTEGRAL in the 511~keV line and the ortho-positronium continuum.

We content ourselves to calculate the total secondary emission of electrons and positrons in the course of their propagation, the only free parameter being the mass of the LDM particle. Energy losses are central to the problem, since all cross-sections of interest depend on energy, except internal Bremsstrahlung associated to the pair production. We then combine the resulting global theoretical spectra with our morphological assumptions to confront them to observations.  The main difficulty is not theoretical, but observational: it will be to separate the bulge emission in the continuum from that of the disk, due to cosmic rays.

\section{Overview}
When colliding in the central part of the Galaxy, LDM particles $\chi$, with mass $m_\chi < 200\mbox{~MeV}$, are expected to annihilate into $\mathrm{e^{-}-e^{+}}$ pairs. The total energy imparted to the pair is equal to twice this mass $m_\chi$, still unknown. Internal Bremsstrahlung in the course of the pair production \cite{Beacom2005a,Boehm2006} produces a continuous spectrum independent of the density of the surrounding medium. Due to the mass limitations, both electrons and positrons are intrinsically of modest energy and their energy losses (proportional to the interstellar density) are mainly due to Coulomb interactions with the surrounding medium. In the course of their propagation, they mainly radiate through the (external) Bremsstrahlung process.
 Moreover, the released positrons are deemed to annihilate with ambient electrons, producing a definite $\gamma$-ray signature (511 keV line plus a low energy continuum). Positron annihilation can occur in flight~\cite{Beacom2005b} ---in 26\% of the cases at most, for $m_\chi = \mbox{200~MeV}$--- or at rest ---mostly through positronium formation. Thus, through the combination of these various mechanisms, a specific $\gamma$-ray spectrum, with a discrete and continuous component, is produced by positrons between their creation and their annihilation.
 
 Our aim is to set an upper limit to the mass of the LDM particle self-consistently, by computing this spectrum and confronting it to the observations in the relevant energy range, \emph{i.e.} between $\sim\mbox{10 keV}$ and 200~MeV. From the particle physics point of view this new constraint, combined with distinct and independent ones coming from big-bang nucleosynthesis \cite{Serpico2004}, type Ia supernov\ae{}, the extra-Galactic hard X-ray and soft $\gamma$-ray background \cite{Ahn2005, Rasera2006}, should help finding a candidate with the appropriate mass and annihilation cross-section properties.

\section{Model\label{sec:model}}
We develop a simple steady-state model adequate for the low energies we are considering, making the reasonable assumption that electrons and positrons below 200 MeV are efficiently confined to the GB, where they annihilate, due to their small gyroradius ---a common hypothesis in this field. We also consider that the only positrons present in the GB are produced in situ, at variance with Prantzos \cite{Prantzos2006}, who speculates  that there could be a  transfer of positrons from the disk to the bulge. This daring possibility is very hypothetical since the magnetic field of the Galaxy at large scales and the propagation of low energy positrons are poorly known, as admitted by Prantzos himself.
 Our problem involves simple physics, repeatedly employed for instance in solar flares physics (\cite{Murphy2005} and references therein) and Galactic physics \cite{Skibo1996}, and various cross-sections that should be carefully chosen. 

\subsection{Steady-state equilibrium}
We consider mono-energetic positrons of total energy $E_{\mathrm{inj}}$ injected in the GB at a rate $R_{\mathrm{inj}}$.
As they travel through the interstellar medium, they loose energy to the ambient medium and can undergo in-flight annihilation.
Supposing the positrons are confined in the bulge and neglecting their spatial distribution, the differential number $N(E)$ of positrons with energy $E$ in steady-state obeys to the diffusion-loss equation
 \begin{equation}
      \frac{d}{dE}\left(-N\frac{dE}{dt}\right) = -Q(E) \,,
 \label{eq:diffusion-loss}     
 \end{equation}
where $\frac{dE}{dt}$ denotes the mean energy loss rate of positrons and the source term
 $Q(E) = R_{\mathrm{inj}} \delta(E-E_{\mathrm{inj}}) - Q_{IA}(E)$ aggregates the mono-energetic injection and the in-flight annihilation, which is a function of energy:
 \begin{equation}
     Q_{IA}(E)= n_e \sigma_{IA}(E)v(E)N(E) \,,
 \label{eq:qia}
 \end{equation}    
with $n_e$ the electron density and $\sigma_{IA}$ the in-flight annihilation cross-section.
Introducing $g(E) = \frac{n_e \sigma_{IA}(E)v(E)}{-dE/dt}$ and $G(E) = \int_{E_{\mathrm{inj}}}^E g(E')dE' = \log{p(E)}$, where $p(E)$ is the probability for a positron injected at $E_{\mathrm{inj}}$ to survive until energy $E$, the number of positrons in the bulge follows from Eq.~\ref{eq:diffusion-loss} and \ref{eq:qia}:
	\begin{equation}
\begin{array}{l@{\quad}l}
N(E) = \frac{R_{\mathrm{inj}}}{-dE/dt} e^{G(E)}	&	(mc^2<E<E_{\mathrm{inj}})\,.\\
\end{array}
	\label{eq:Npos}
 \end{equation}
We can then obtain the total rate of positrons undergoing in-flight annihilation
	\begin{equation}
     R_{IA} = \int_{mc^2}^{E_{\mathrm{inj}}}  Q_{IA}(E)dE = R_{\mathrm{inj}} [1 - e^{G(mc^2)}]~\mathrm{e^+\,s^{-1}}\,,
 \label{eq:ria}
 \end{equation}
and deduce the rate of annihilation at rest, assuming it equals the rate of positrons reaching complete rest:
\begin{equation}
R_{A} = R_{\mathrm{inj}} e^{G(mc^2)}~\mathrm{e^+\,s^{-1}}.
\label{eq:Ra}
\end{equation}

\subsection{Energy losses}

The total energy lost by positrons during thermalization goes into various interactions -- excitation or ionization, Bremsstrahlung, inverse Compton and synchrotron effects. As their importance varies with the positron's energy, it is mandatory to use refined energy loss rates. For the range of injection energies we are interested in (up to 200~MeV at most), we can limit ourselves to ionization and Bremsstrahlung.

We stress that the ionization degree $x_i = n_{H_{II}}/n_H$  in the region of propagation, where $n_H$ is the hydrogen number density, is of prime importance for the estimate of Coulomb losses \cite{Soutoul1989}, since at a given energy they are several times weaker in a neutral medium than in a plasma.

\begin{figure}
   \centering
   \includegraphics[width=0.4\textwidth]{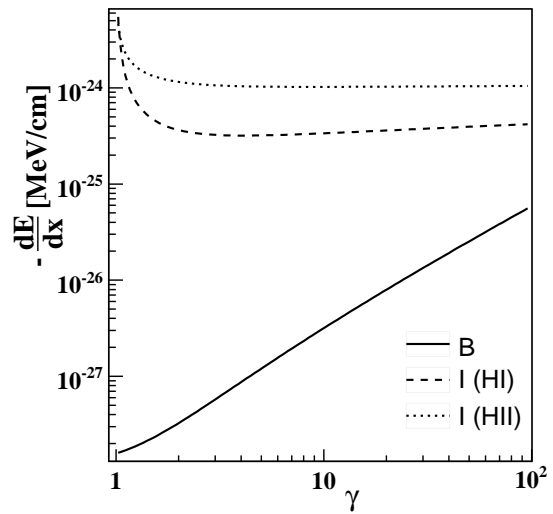}
   \caption{Energy losses of positrons in a 51\% ionized medium ($n_H = 0.1\mathrm{\,cm^{-3}}$): energy loss rate per unit path length $\frac{dE}{dx}$ as a function of the positron's Lorentz factor $\gamma$. Losses through Coulomb collisions within an ionized medium (dotted) prevail on ionization of the neutral phase (dashed), but both greatly exceed Bremsstrahlung losses (plain).}
   \label{fig:losses}
\end{figure}

Thus, the  total loss rate sums up three terms:
	\begin{equation}
     \frac{dE}{dt} = \frac{dE^B}{dt}(n_H) + \frac{dE^{I, HI}}{dt}(n_H, x_i) + \frac{dE^{I, HII}}{dt}(n_H, x_i)\,,
 \label{eq:losses}
 \end{equation}
where the energy lost by a positron due to ionization of neutral matter is taken from \cite{Longair1992} while the loss rate for excitation of the ionized component follows \cite{Aharonian1981a}. The energy loss caused by Bremsstrahlung interactions is given by Eq.~4BN in \cite{Koch1959}. The relative amplitude of each component is illustrated in Fig.~\ref{fig:losses}, which shows energy losses $\frac{dE}{dx}$ per unit path length.

\subsection{Radiation spectra}
\subsubsection{Positron radiation}
The total radiation spectrum due to positrons is the sum of five components: the 511~keV line from $\mathrm{e^+\mbox{-}e^-}$ annihilation at rest and from para-positronium decay into two gamma rays
 (i), the ortho-positronium spectrum from its $3\gamma$ decay (ii), the continuous spectrum from $\mathrm{e^+\mbox{-}e^-}$ in-flight annihilation into $2\gamma$ (iii), the radiation spectra associated with Bremsstrahlung energy losses (iv) and with the internal Bremsstrahlung taking place in the annihilation process of the dark matter particle (v).

The flux from $2\gamma$ annihilation at rest $F^{2\gamma}$ is tuned via the injection rate $R_{\mathrm{inj}}$ to the value of $(1.07\pm{}0.03)\,{10}^{-3}~\textrm{ph}/\textrm{cm}^2/\textrm{s}$ deduced from INTEGRAL observations \cite{Jean2006} :
	\begin{equation}
F^{2\gamma} = 2\frac{(1-f_{Ps})+f_{Ps}/4}{4\pi d_{GC}^2} R_A,
	\label{eq:F2g}
 \end{equation}
 using the associated positronium fraction value \mbox{$f_{Ps} = 0.967\pm{}0.022$} \cite{Jean2006} and the relationship (\ref{eq:Ra}) between the initial positron injection rate and the rate $R_A$ of positron annihilation at rest.

The $3\gamma$ ortho-positronium decay spectrum follows

\begin{equation}
	\frac{dF^{3\gamma}}{dE_\gamma} = 3\frac{3f_{Ps}/4}{4\pi d_{GC}^2} R_A \frac{df_{3\gamma}}{dE_\gamma}(E_\gamma)\,,
	\label{eq:dF3gdE}
\end{equation}
where the spectral distribution $\frac{df_{3\gamma}}{dE_\gamma}$ was derived by Ore \& Powell \cite{Ore1949} in \citeyear{Ore1949}.

\begin{figure}
   \centering
   \includegraphics[width=0.4\textwidth]{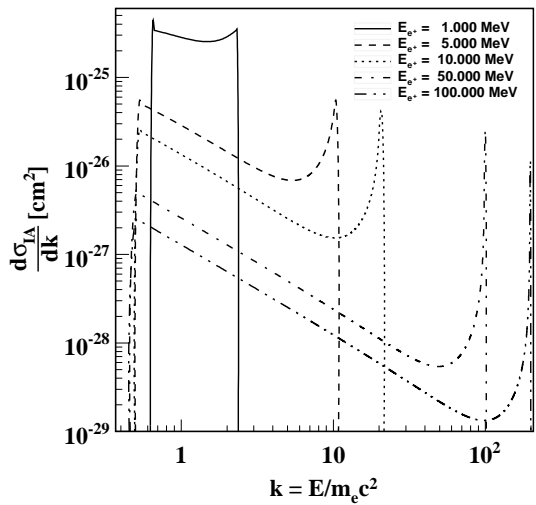}
      \caption{Differential in-flight annihilation cross-section --given for instance by \citet[Eq.~5,~misprint~corrected]{Aharonian1981b}-- as a function of the reduced photon energy, for various positron energies. Still energetic positrons annihilate with thermal electrons into two photons that share the total available energy, with a minimum energy of $m_ec^2/2$ by photon.}
   \label{fig:dSdkIA}
\end{figure}

Convolving the differential in-flight annihilation cross-section (Fig.~\ref{fig:dSdkIA}) with the positron distribution yields the in-flight annihilation component:
\begin{equation}
	\frac{dF^{IA}}{dE_\gamma} = \frac{1}{4\pi d_{GC}^2}\int_{Emin(E_\gamma)}^{E_{\mathrm{inj}}}{n_e \frac{d\sigma^{IA}}{dE_\gamma}(E_\gamma,E)v(E)N(E)dE}\,.
	\label{eq:dFiadE}
\end{equation}
The external Bremsstrahlung spectrum is derived in a similar way using Koch \& Motz~\cite{Koch1959}'s  cross-section 3BN.
Dependence on the electron number density $n_e$ cancels out as both the differential emissivity $n_e\frac{d\sigma}{dE_\gamma}v$ and the energy losses in the positron spectrum $N(E)$ (see Eq.~\ref{eq:Npos}) are directly proportional to $n_e$. Finally, the internal Bremsstrahlung spectrum depends only on the positron injection rate:
\begin{equation}
	\frac{dF^{IB}}{dE_\gamma} = \frac{1}{4\pi d_{GC}^2}\frac{1}{\sigma_{tot}}\frac{d\sigma^{IB}}{dE_\gamma}\,
	\label{eq:dFibdE}
\end{equation}
 where we use the differential cross-section first derived by Beacom et al. \cite{Beacom2005a}. Another cross-section has been derived by Boehm \& Uwer \cite{Boehm2006} since then, which appears to be larger.

\subsubsection{Electron radiation.}
In addition to the five positron radiation components comes the Bremsstrahlung spectrum from the mono-energetic electrons injected together with positrons during the dark matter annihilation process, whose equilibrium spectral distribution follows 
\begin{equation}
	\begin{array}{l@{\quad}l}
	N_{e^-}(E) = \frac{R_{\mathrm{inj}}}{-dE/dt}	&	(mc^2<E<E_{\mathrm{inj}})\\
	\end{array}
	.
	\label{eq:Nel}
\end{equation}

\subsection{Discussion}
\begin{figure}
   \centering
   \includegraphics[width=0.4\textwidth]{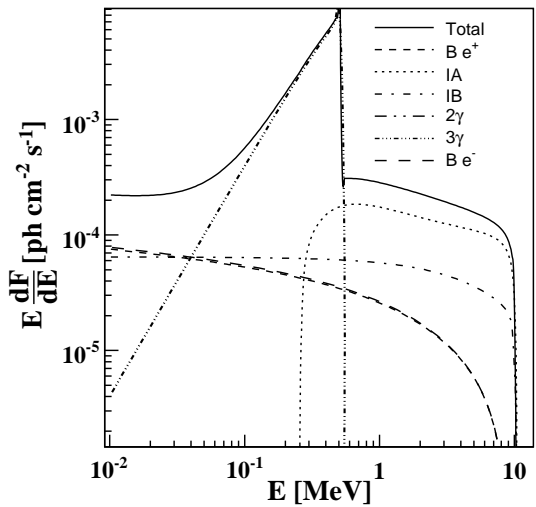}
   \caption{Light dark matter ($E_{\mathrm{inj}}$=10~MeV) radiation spectra: positron annihilation spectra (in-flight:~IA; 2$\gamma$ at rest; 3$\gamma$ at rest), positron energy loss spectra (internal Bremsstrahlung: IB; Bremsstrahlung: B $e^+$), electron Bremsstrahlung spectrum (B $e^-$) and total spectrum (plain).}
   \label{fig:spetot}
\end{figure}

\begin{figure}
   \centering
   \includegraphics[width=0.4\textwidth]{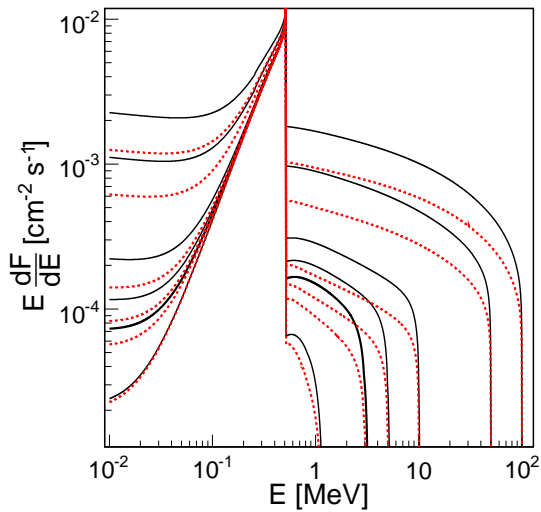}
      \caption{Total dark matter radiation spectrum for increasing injection energies $E_{\mathrm{inj}}$ (1, 3, 5, 10, 50 and 100~MeV). In a partially ionized medium (red dotted curves), higher energy losses through ionization lead to lower external Bremsstrahlung fluxes than in a completely neutral medium (black solid curves).}
   \label{fig:spetot2}
\end{figure}

Fig.~\ref{fig:spetot} displays the various components of the radiation emitted by positrons and electrons from dark matter annihilation for a specific mono-energetic injection. While external Bremsstrahlung becomes the main component at very low energies, it is negligible compared to both in flight annihilation and internal Bremsstrahlung above 1~MeV, which we will show to be the most constraining energy domain. Fig.~\ref{fig:spetot2} shows the evolution of the total spectrum with increasing injection energies and the influence of the ionization fraction.

Using a multi-component model of the interstellar medium (molecular, cold, warm neutral, warm ionized, hot), \citeauthor{Jean2006} \cite{Jean2006} showed, through their analysis of the 511~keV line and the ortho-positronium spectrum, that $51^{+3}_{-2}\%$ of the rest annihilation takes place in the warm ionized phase of the ISM. We can assume that the mean degree of ionization of the medium where the thermalizing positrons propagate ranges between 0 and 51\%. We consider both extreme cases; the latter one enables us to release constraints on the mass of the dark matter particle.

\section{Confrontation to observations}
In section \ref{sec:model}, we computed the total diffuse $\gamma$-ray flux caused by the production, interactions and annihilation of positrons and electrons coming from the annihilation of a LDM particle (hereafter the LDM flux).  We now would like to use observations by INTEGRAL (ISGRI and SPI) and CGRO (COMPTEL and EGRET) to derive some constraints on the mass of this particle. The computed flux stands for the GB, in which we assume that positrons (and associated electrons) are produced and trapped.

Therefore, we would like to show, for each possible injection energy of the positrons :
\begin{itemize}
	\item[(i)] whether the available data should and do display an excess in the bulge region with respect to the surrounding Galactic plane;
	\item[(ii)] whether the total predicted flux exceeds the observed one, taking other gamma-ray diffuse components into account.
The latter include extra-Galactic diffuse emission, diffuse emission from the interaction of Galactic cosmic rays (CRs) with the interstellar medium and a component due to ---still--- unresolved point sources.
\end{itemize}

\subsection{Observational data}
We consider the available measurements for diffuse emission in the 20~keV to 200~MeV energy range (data points in Fig.~\ref{fig:broadband} and \ref{fig:pow}).

Analyzing the INTEGRAL/IBIS/ISGRI data, Krivonos et al. \cite{Krivonos2006} computed the Galactic ridge X-ray spectrum in the 17 to 200~keV range using the imager IBIS as a collimated telescope. They derived the residual emission from the total field of view remaining after subtraction of the instrumental background and of resolved point sources. Considering the $\sim{}$19\degrees{} width of the 50\% coded field of view, we will be able to compare the fluxes they derived to models of a 20\degrees{} wide region around the Galactic center. Below $\sim{}$60~keV, there remains a residual flux unexplained by resolved point sources. Whether this flux is due to truly diffuse emission or can be attributed to a population of unresolved sources, like Galactic pulsars, is still matter of debate. Above 60~keV, no residual flux is detected but 2$\sigma$ upper limits on the possible unresolved emission have been set for three energy bands up to $\sim{}190$~keV.

For the diffuse emission observed by the INTEGRAL/SPI spectrometer, we use the spectrum obtained by Strong et al.~\cite{Strong2005} in the 18~keV to 1.018~MeV energy range. Note that it has also been studied independently by \citeauthor{Bouchet2005} \cite{Bouchet2005} who derived similar conclusions. \citeauthor{Strong2005} proceeded by fitting the instrument's data with a multi-component imaging model including point sources detected by the spectrometer and a set of ten maps with various diffuse morphologies. The sum of the ten diffuse components represents the map of the emission seen by SPI as diffuse. Strong et al. published (\cite{Strong2005}, Tab.~B.4) the fluxes of this map integrated over a 20\degrees{}$\times$20\degrees{} region at the Galactic center, shown in Fig.~\ref{fig:broadband} and \ref{fig:pow}. The associated uncertainties do not include systematic effects. We will compare these fluxes to a model of emission in a region of the same size, as a more detailed morphology of the SPI emission inside this region is not available yet apart from the morphology studies of the para-positronium and 511~keV annihilation components.
We notice a discrepancy between the SPI and ISGRI diffuse spectra. This gap could be explained by the imager's better ability to detect faint point sources, considered as diffuse by SPI. However, there might be a need for a more profound analysis.

For CGRO/COMPTEL, we will start from the intensity skymaps of diffuse emission produced by \citeauthor{Strong1999} \cite{Strong1999} in the 1-3, 3-10 and 10-30~MeV energy bands. These maps are available with a resolution of 1\degrees{}. After subtracting a zero level deduced from high latitudes, we compute total fluxes in regions of variable sizes. Data points in Fig.~\ref{fig:broadband} and \ref{fig:pow} correspond to a 20\degrees{}$\times$20\degrees{} region, for mere compatibility with the ISGRI and SPI observations.
Uncertainties on these maps are largely dominated by systematic effects; an estimation based on local fluctuations at high latitudes leads to uncertainties of at least 30\%  \cite{StrongP,Strong1999}. Hence, for all sizes of our integration region, we consider the COMPTEL flux uncertainty to be 30\%.

We also consider CGRO/EGRET $\gamma$-ray measurements \cite{Hunter1997} above 30~MeV~; the fluxes of data points in Fig.~\ref{fig:broadband} and \ref{fig:pow} are obtained by subtracting the isotropic extra-Galactic component to longitude profiles and integrating them on our region of study; associated systematic uncertainties amount to $\sim$15 to 20\%.

\subsection{LDM morphology hypothesis}
Our model does not include spatial diffusion but simply assumes that positrons injected in the Galactic bulge remain confined to it; therefore, we have no detailed information on the morphology of the LDM emission. However, analysis of the SPI data showed that the 511~keV component is compatible with a 8\degrees{}~FWHM Gaussian spatial distribution \cite{Knodlseder2005} and that the ortho-positronium continuum component had a similar spatial distribution \cite{Weidenspointner2006}. Thus, we can safely assume that the rest of the LDM emission, \emph{e.g.} radiation emitted during the injection and thermalization phases, comes from the same region or a smaller one.

When considering a 20\degrees{}$\times$20\degrees{} region, we do not need to make a strong assumption on the morphology of in-flight annihilation and we can use the total LDM flux computed in section \ref{sec:model}. Conversely, when considering a smaller region, we assume an LDM morphology following a Gaussian distribution with an 8\degrees{} FWHM and we scale the total LDM flux by the appropriate fraction corresponding to the solid angle subtended by this region.

In a first step, we use the 20\degrees{}$\times$20\degrees{} region which is the smallest one for which data for all four instruments are available. In a second step, we only use the COMPTEL data, which bring the most constraining information, and we consider circular regions of 5, 8 and 14.6\degrees{} of diameter corresponding to regions covering respectively 24, 50 and 90\% of the LDM flux. Such smaller regions include less diffuse background from cosmic-ray interaction and enable to enhance the signal to background ratio, leading to more stringent constraints.

\subsection{Full model approach}
\label{sec:totalbulge}

\begin{figure}[!htbp]
   \centering
	\includegraphics[width=0.5\textwidth]{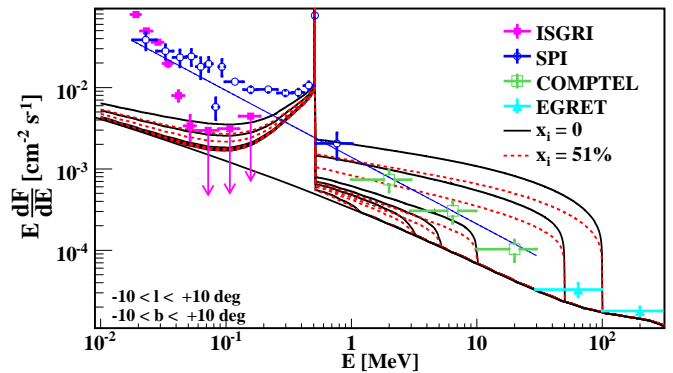}   
   \caption{Comparison between the total diffuse spectrum expected from a 20\degrees{} wide region at the Galactic center and the diffuse fluxes measured by IBIS (Krivonos et al. 2006~\cite{Krivonos2006}; magenta), SPI (Strong et al. 2005~\cite{Strong2005}; blue), COMPTEL (Strong et al. 1999~\cite{Strong1999}; green) and EGRET (Hunter et al. 1997~\cite{Hunter1997}; cyan). The model curves are the sum of the CR interactions component computed by \citeauthor{Strong2005}~\cite{Strong2005} (lower black line), and the dark matter related component, computed in \S~\ref{sec:model} for injection energies $E_{\mathrm{inj}}$ = 1, 3, 5, 10, 50 and 100~MeV, in a neutral (black solid curves) or 51\% ionized medium (red dashed curves).}
   \label{fig:broadband}
\end{figure}

The main problem when comparing LDM theoretical emission to the data is to estimate the other diffuse components, mainly the emission due to interactions of cosmic rays with the interstellar medium. In this paragraph, we adopt the approach consisting in reproducing the observations using a complete model.

We consider the \texttt{GALPROP} code developed by Strong et al. \cite{Strong2004b,Strong2005} to model the cosmic-ray propagation and the associated $\gamma$-ray production. The total CR gamma-ray emission is composed of three main components~: Bremsstrahlung scattering, the inverse Compton effect, and pion decays. The relative importance of these components varies with energy.

Fig.~\ref{fig:broadband} shows, for a 20\degrees{}$\times$20\degrees{} region, the spectra of the total diffuse emission expected. They sum up the radiation from CRs, obtained by integration of the \texttt{GALPROP} (version 600203a) skymaps, and the total LDM $\gamma$-ray flux, for various positron injection energies reflecting the LDM particle mass. For each injection energy, both the completely neutral propagation medium case and the 51\% ionized medium case are illustrated.

The \texttt{GALPROP} model reliably reproduces the measured diffuse spectrum over a wide energy range, from tens of MeV to hundreds of GeV. In particular, it reproduces the EGRET measurements satisfactorily, mostly through the pion decay bump.
On the contrary, at lower energies, this CR model fails to account by itself for all of the detected diffuse emission.
Assuming that this model still correctly predicts the diffuse emission due to cosmic-ray interactions below 30~MeV, this leaves room for diffuse emission from a different origin or for unresolved point sources.
In this section, we tentatively adopt the hypothesis that all of the discrepancy in the bulge between the data and the \texttt{GALPROP} model is due to the sole LDM emission.

Then, Fig.~\ref{fig:broadband} permits us to derive for each instrument constraints on the maximum mass of the LDM particle. With a 95\% confidence level, in the more stringent case of a neutral medium, we exclude LDM particles with masses greater than $\sim$70~$\mathrm{MeV/c^2}$ using the ISGRI upper limits, 200~$\mathrm{MeV/c^2}$ using the SPI flux above the 511 keV line, 25~$\mathrm{MeV/c^2}$ using the 10-30~MeV COMPTEL flux. For EGRET, the cosmic-ray interaction models suffice to explain the measurements; no additional diffuse component such as in-flight annihilation or internal Bremsstrahlung is required, which sets the maximum LDM particle mass to the lower energy boundary of the instrument of 30~MeV.

\begin{figure*}[htbp]		
		\includegraphics[height=0.5\textwidth]{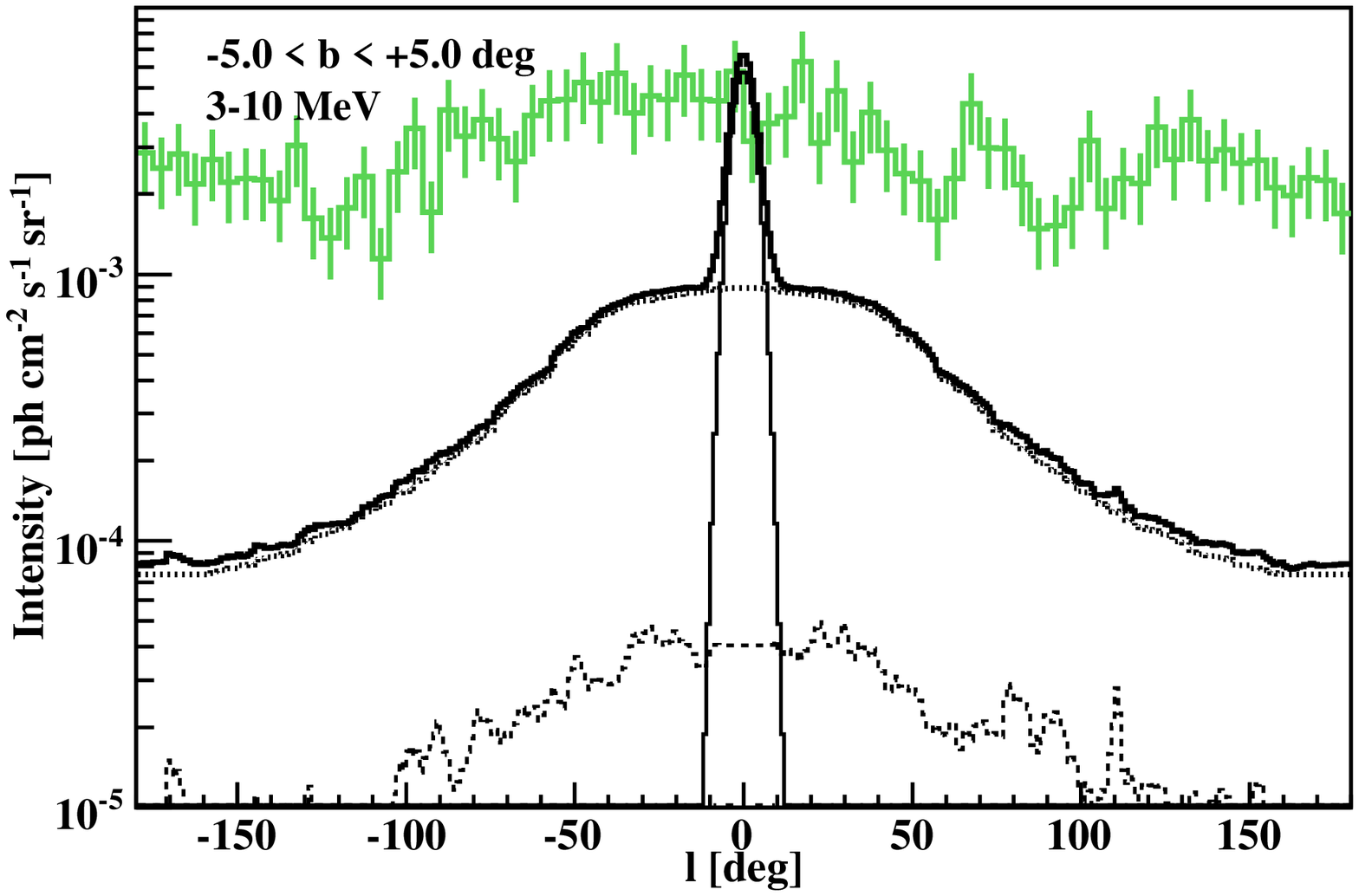}
		\includegraphics[height=0.5\textwidth]{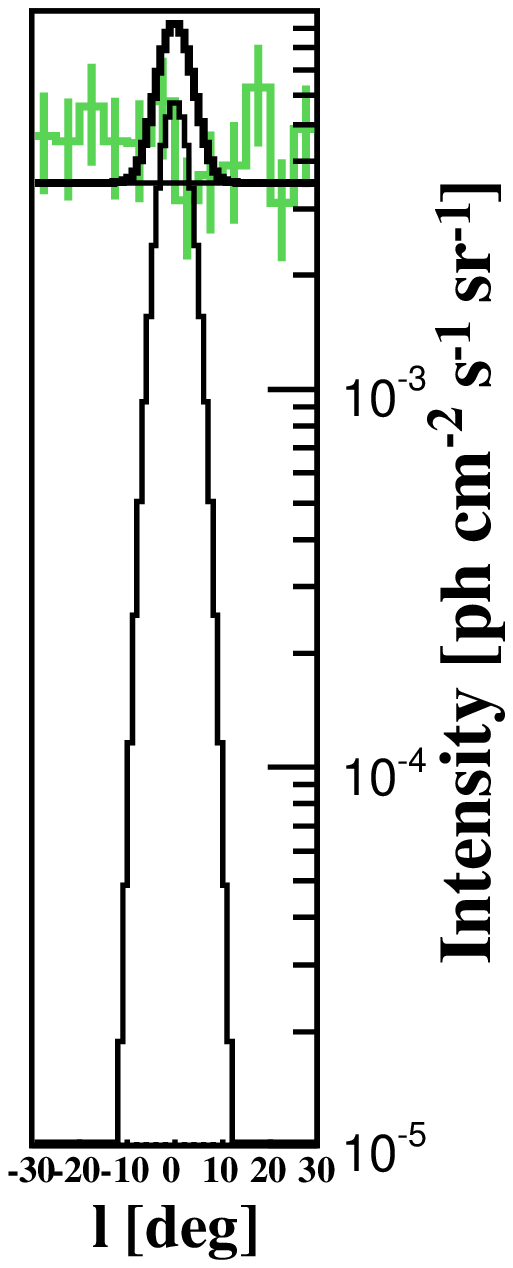} 
		\caption{Longitude profile of the 3-10 MeV diffuse flux measured by COMPTEL in a latitude band of $\pm$5\degrees{} (histogram).\\ 	Left~: the data is compared to a global model ---thick line--- including the GALPROP model of diffuse emission from cosmic-ray interactions, composed of a dominating inverse Compton component ---dotted line--- and a smaller Bremsstrahlung component ---dashed line---, to which we added the model of LDM emission ---thin solid line---, computed here for an injection energy $E_{\mathrm{inj}}$ of 10~MeV and assumed to have a Gaussian morphology with 8\degrees{} FWHM . As the discrepancy between the COMPTEL data and the GALPROP model concerns high longitudes as well as the Galactic center, trying to explain all of the Galactic bulge gap through the LDM emission is not reasonable because the outer-bulge disparity would remain unexplained.\\
	Right : thus, in \S\ref{sec:morph} we dismiss the idea of modeling the observations completely; we acknowledge the relatively flat profile of the COMPTEL data in the inner disk region and use the mean intensity as a baseline, requiring that an eventual excess caused in the Galactic bulge by LDM radiation remains within data uncertainties.\label{fig:profiles}}
\end{figure*}

But the hypothesis adopted that all of the discrepancy between the data and the \texttt{GALPROP} model in the Galactic bulge is due to the sole LDM emission suffers from the fact that the same discrepancy can also be noticed in the Galactic disk, as depicted in Fig.~\ref{fig:profiles}. The longitude profile of the COMPTEL flux in the 3-10~MeV band is rather flat, shows no excess in the bulge region that could be identified as a signature of LDM radiation. In order to fill the gap between the data and the \texttt{GALPROP} model in the bulge with LDM radiation, we would need to simultaneously explain the outer disk deficit, by adding another model component from a totally different origin which would be present only outside the Galactic bulge region. Such an hypothesis seems unrealistic.

Additional tuning of the \texttt{GALPROP} model permits to enhance the predicted CR flux in the COMPTEL energy range, notably by using a less conventional model with a rapid upturn in the CR electron spectrum \cite{Moskalenko1999,Strong2000}. However, this idea causes energetic problems and has no natural motivation~: while the origin of the discrepancy with the data is still debated, a population of unresolved sources is more likely \cite{Strong2000}. Moreover, such a model would explain the COMPTEL observations over the whole longitude range. Thus, we should derive our constraints simply through the absence of any excess in the bulge with respect to the disk in the COMPTEL data.

\subsection{Empirical approach}
\label{sec:morph}
In this paragraph, we drop the idea of using a full model approach which, as discussed previously, has the drawback of leaving completely out of consideration the fact that, outside the energy range of the positronium \cite{Weidenspointner2006} and the 511~keV line \cite{Knodlseder2005}, no continuum excess has been detected in the bulge with respect to the surrounding Galactic plane.

Therefore, taking note of the more disc-like than bulge-like morphology suggested by COMPTEL skymaps and profiles above 1~MeV, we assume that most of the emission detected by both COMPTEL and SPI have their origin in either unresolved faint point sources or diffuse mechanisms other than those related to LDM particles. Assuming a relative uniformity of this $\gamma$-ray background along the Galactic plane, we derive the mean COMPTEL intensity in a wide region ($|l|<30^\circ, |b|<5^\circ$) and use it as an estimate of background for smaller regions. An additional signal caused by LDM particles is then searched as an excess over this background which should remain within data uncertainties. This approach, also used by \citeauthor{Beacom2005b}~\cite{Beacom2005b}, is illustrated in the right panel of Fig.~\ref{fig:profiles}. 

\begin{figure*}[!bhpt]
   \centering
   \includegraphics[width=\textwidth]{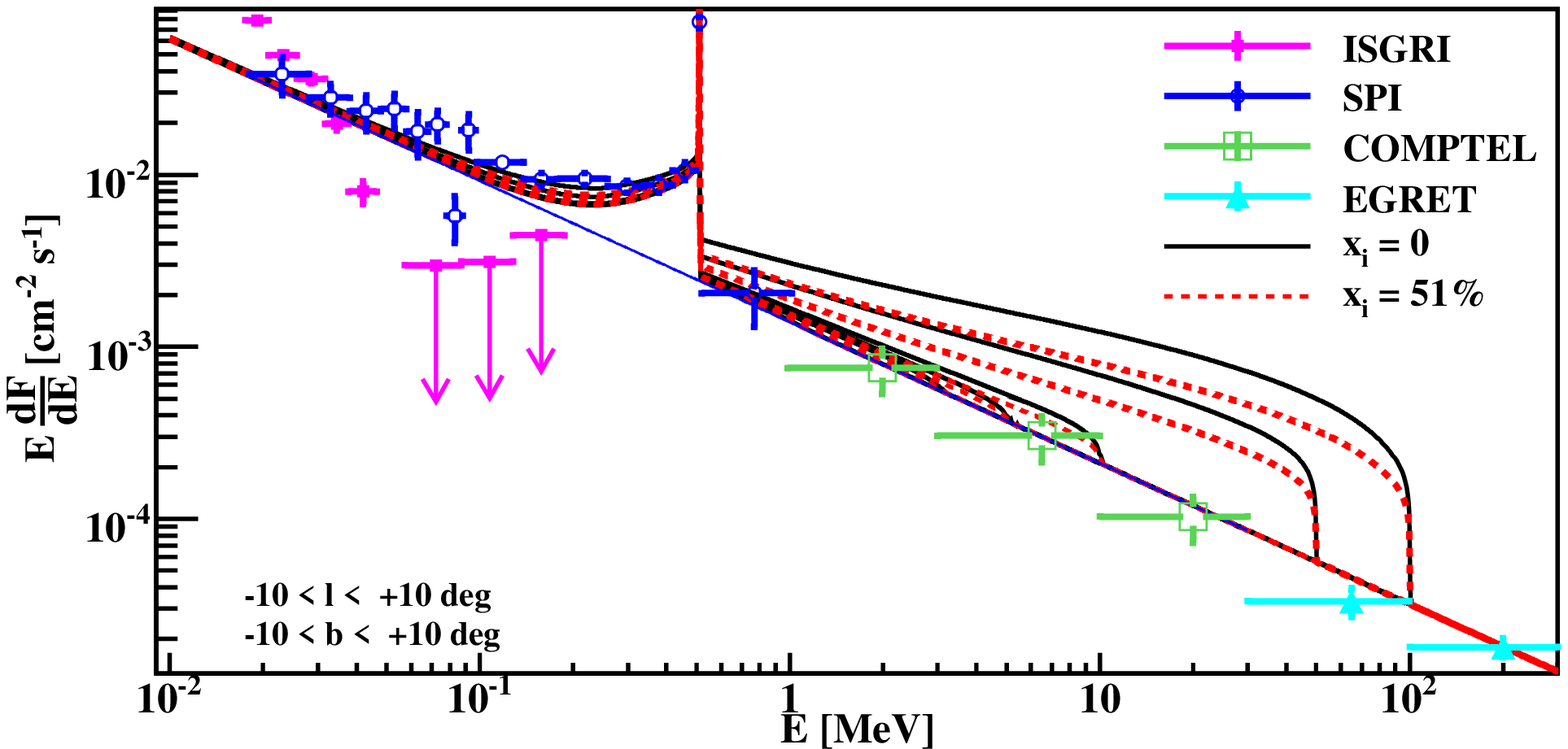}
   \caption{Comparison between the total diffuse spectrum expected from a 20\degrees{} wide region at the Galactic center and the diffuse fluxes measured by IBIS (magenta), SPI (blue) and COMPTEL (green). Instead of taking the predictions from cosmic-ray interaction codes as a base level for our LDM model like in Fig.~\ref{fig:broadband}, we now use a powerlaw (blue straight line) $\gamma$-ray background ---already displayed in Fig.~\ref{fig:broadband}--- compatible with both SPI and COMPTEL data.}
   \label{fig:pow}
\end{figure*}

\begin{figure}[!htbp]
	\includegraphics[width=0.4\textwidth]{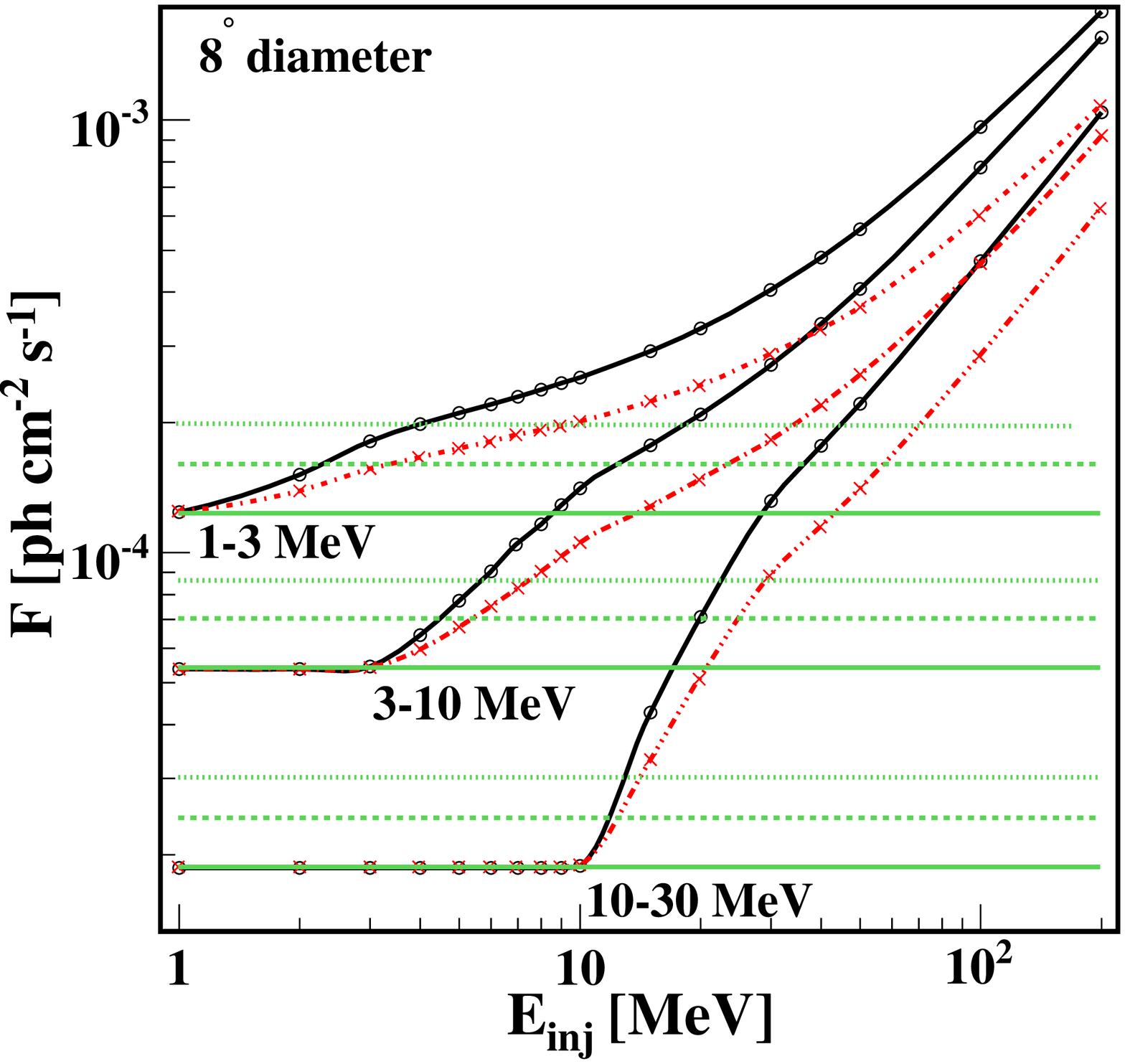}
	\caption{Total integrated flux in the 1-3, 3-10 and 10-30~MeV COMPTEL energy bands for an 8\degrees{} diameter region around the Galactic center. Horizontal lines show the flux ---solid lines--- derived from the mean COMPTEL intensity in a wider fraction of the disk ($|l| < 30^\circ $, $|b|<5^\circ $) and its 1$\sigma$ ---dashed--- and 2$\sigma$ ---dotted--- upper values. The sum of the COMPTEL flux and of the LDM flux is shown as a function of the positron injection energy $E_{\mathrm{inj}}$, for a completely neutral propagation medium --- circled solid lines--- and for 51\% ionized medium --- crossed dot-dashed lines.
	It should be noticed that combining the three COMPTEL energy bands instead of considering them independently would bring a  sensitivity improvement and lead to more severe exclusions; for instance, injection at $E_{inj}$=10~MeV is excluded by the 1-3 and 3-10~MeV energy bands at 3.6$\sigma$ and 5.6$\sigma$ respectively ($x_i$ = 0), while the combination of both bands leads to a 6.9$\sigma$ exclusion.}
	\label{fig:fluxtot}
\end{figure}

Fig.~\ref{fig:pow} displays total diffuse radiation spectra. The LDM radiation spectra, for various positron injection energies, have been added to a powerlaw spectrum based on the data itself, at variance with Fig.~\ref{fig:broadband} where the \texttt{GALPROP} cosmic-ray induced spectrum was used as a baseline. We took the powerlaw derived for the complete SPI unresolved spectrum by Strong et al. \cite{Strong2005},
\begin{equation}
	\frac{dF}{dE} = (1.39\pm{}0.27)\,10^{-3} (\frac{E}{\mathrm{MeV}})^{-1.82\pm{}0.08}\,\mathrm{cm^{-2}\,s^{-1}\,MeV^{-1}},
	\label{eq:strongpowerlaw}
\end{equation}
 which is remarkably compatible with the COMPTEL data points.
This figure still covers a 20\degrees{}$\times$20\degrees{} region to display the ISGRI and SPI data points. However, in order to derive constraints, different sizes of the integration region will be taken into account.

For this purpose, we consider the 1-3, 3-10 and 10-30~MeV COMPTEL energy bands, which are the most constraining. Fig.~\ref{fig:fluxtot} illustrates the constraints on the LDM particle mass obtained when using an 8\degrees{} diameter region. It displays, as a function of the positron injection energy and for each energy band, the sum of the total COMPTEL flux and of the LDM radiation flux, scaled by a factor 50\% corresponding to the fraction of the total LDM radiation coming from this region under our assumption that LDM radiation follows a spatial distribution identical to that of the observed 511~keV line, \emph{i.e.} an 8\degrees{} FWHM Gaussian.
The 1$\sigma$ and 2$\sigma$ upper values on the COMPTEL fluxes are also shown.
By excluding injection energies yielding a total COMPTEL~+~LDM flux exceeding the 2$\sigma$ COMPTEL upper values, we derive the maximum mass of the LDM particle at a 95\% confidence level. The results are given in Tab.~\ref{tab:constraints}, together with constraints obtained by considering smaller or larger regions, for both the case of a completely neutral propagation medium and a medium with an ionization fraction of 51\%. 

In the case of a neutral medium, we set a maximum of $\sim{}4\,\mathrm{MeV}$ on the positron injection energy with the integration region covering the FWHM of the emission.
When taking a 5\degrees{} diameter, corresponding to 25\% of the LDM flux, we maximize the signal to background ratio and obtain results similar to those of \citeauthor{Beacom2005b} \cite{Beacom2005b}~: a maximum mass of $\sim{}3\,\mathrm{MeV/c^2}$ for the light dark matter particle. The precise limit depends on the not so well known uncertainties on the COMPTEL measurements.

	The value of the fraction of ionized phase in the medium the positrons propagate through has a small but not negligible influence on our LDM particle mass constraints. By assuming an ionization fraction of 51\%, those constraints can be released by a factor of approximately two (details in Tab.~\ref{tab:constraints}).

\begin{table}[!hbtp]
	\centering
		\begin{tabular}[t]{|c|c||c|c|c|}
		\hline
		\multirow{2}{1cm}{Size [\degrees{}]} & \multirow{2}{1cm}{$x_i$ [\%]} & \multicolumn{3}{|c|}{Energy band} \\
		\cline{3-5}
		&	 & 1-3~MeV & 3-10~MeV & 10-30~MeV \\
		\hline
		\hline
			\multirow{2}{1cm}{\diameter~5} &  0    & \textbf{3} & 5.5 & 12 \\
		\cline{2-5}			
																		 &  51 & 7 & \textbf{6.5} & 13 \\
		\hline																		 
			\multirow{2}{1cm}{\diameter~8} &  0    & \textbf{4} & 6 & 12\\
		\cline{2-5}			
																		 &  51 &	9 & 	\textbf{7.5} & 13.5 \\	
		\hline
			\multirow{2}{1cm}{\diameter~14.6} &  0 & 10 & \textbf{8} & 14.5\\
		\cline{2-5}				
																		 &  51 &	23 & \textbf{12} & 17 \\	
		\hline																		 
			\multirow{2}{1cm}{20$\times{}$20} &  0  & 28  & \textbf{16} & 18.5  \\
		\cline{2-5}				
																		 &  51 &	60 & 27 & \textbf{23} \\ 												 
		\hline
		\end{tabular}
	\caption{Constraints from COMPTEL data on the maximum positron injection energy $E_{\mathrm{inj}}$ [MeV]. The table gives the 2$\sigma$ (95\% confidence) maximum energies deduced from the 1-3, 3-10 and 10-30~MeV COMPTEL measurements respectively, for four different integration regions (5, 8 and 14.6\degrees{} diameter, 20\degrees{} width), including 24\%, 50\%, 90\% or 99\% of the LDM emission respectively ---under the assumtion that positron in-flight annihilation has a morphology similar to that of annihilation at rest.}
	\label{tab:constraints}
\end{table}

\section{Conclusions}
Combining all the radiation components related to LDM particles ---mostly in-flight annihilation and internal Bremsstrahlung--- we were able to set an upper bound of 3 to 7.5~MeV on the mass of these particles, based mostly on COMPTEL data.

A previous study by \citeauthor{Beacom2005b} \cite{Beacom2005b} had concluded on a 3~$\mathrm{MeV/c^2}$ upper mass. All in all, this work shows essential agreement with their result; no inconsistencies were found when making similar assumptions. More conservative morphology choices lead to slightly larger values but the effect of the ionization fraction of the ISM is more noticeable. Arguably, there seems to be a tension between the lower limit established through the constraint of supernovae explosions \cite{Fayet2006} and the upper limit based on in flight annihilation (\cite{Beacom2005b} and this work), but a definite exclusion of the LDM hypothesis is premature, since the discussion on the uncertainties on the measurements in the COMPTEL energy band, which are of crucial importance, is rather qualitative. To advance in this context, a refined reassessment of both statistical and systematic uncertainties on fluxes in COMPTEL maps would certainly prove to be useful. Before definitely excluding a light dark matter particle of mass less than about 10~$\mathrm{MeV/c^2}$, further observations of the Galactic bulge region with INTEGRAL/SPI are mandatory to get additional morphological information and refined uncertainties on diffuse emission in the 511~keV to 1~MeV energy domain.

We stressed the influence of the ionization degree of the interstellar medium on the amplitude of the in-flight radiation spectrum~: taking into account the ionization fraction of the propagation region could allow values of the mass of the LDM particle larger by up to a factor two. However, additional modeling work is required to ascertain in which phases positrons propagate during their thermalization phase.

The present conclusions were derived under the demanding assumption that light dark matter annihilation alone accounts for all of the 511~keV line. A multi-source scenario would enable us to reduce the amplitude of the LDM $\gamma$-ray continuum and to release once again the limit on the LDM particle mass.

This work focuses on the LDM hypothesis; however, by leaving the internal Bremsstrahlung component aside, the model presented can be applied to the more general case of $\gamma$-ray radiation from any mono-energetic source of positrons.

\paragraph*{Prospects}

In order to further restrict the LDM particle mass range with the INTEGRAL satellite, a refined appraisal of the contribution of faint point sources in the 50-511~keV energy range is required. In this perspective, the new census of sources found by the imager above 100~keV \cite{Bazzano2006} will be paramount to fine-tune ISGRI upper limits on diffuse emission. Because the spectrometer does not have the capacity to resolve as many faint sources, an increased recognition of ISGRI detections in the analysis of SPI data might prove necessary.

Beyond flux considerations, whose precision is hindered by our uncertain knowledge of the radiation from cosmic-ray interactions and unresolved sources, morphology studies will be a key point to find out whether the 511~keV to 8~MeV energy range presents a faint excess in the GB.

In order to take all aspects of LDM into account, the present model could be extended to include spatial diffusion of positrons and electrons during thermalization and model the differences in the morphologies of the LDM emission at various energies.
Furthermore, studying the influence of the in-flight annihilation and internal Bremsstrahlung components on the positronium fraction~\cite{Jean2006} derived from SPI data would be interesting.

\begin{acknowledgments}
The authors thank A.~W.~Strong for making available COMPTEL and SPI results as well as the \texttt{GALPROP} model. We are also grateful to P.~Fayet, F.~A.~Aharonian, B.~Cordier, F.~Lebrun, J.~Paul and P.~Jean for discussions on in-flight annihilation, energy losses and observational data uncertainties, and to the referee for his very careful reading and appropriate suggestions.
\end{acknowledgments}

\bibliography{sizun2006ia_v3}

\end{document}